\pdfoutput=1
\documentclass[conference]{IEEEtran}
\IEEEoverridecommandlockouts
\usepackage{cite}
\usepackage{breqn}
\usepackage{amsmath,amssymb,amsfonts}
\usepackage{graphicx}
\usepackage{textcomp}
\usepackage{xcolor}
\usepackage{acronym}
\usepackage{url}
\usepackage{verbatim}
\usepackage{mathtools}
\usepackage{cleveref}
\usepackage{algorithm} 
\usepackage{algpseudocode} 
\usepackage{amsmath}
\usepackage{amssymb}
\usepackage{amsthm}
\usepackage{bm}

\usepackage{multirow}
\usepackage{tabularx}
\usepackage{booktabs}
\usepackage{rotating}
\usepackage{makecell}
\usepackage{longtable}
\usepackage{array}
\usepackage{lscape}
\usepackage{multirow}
\usepackage{float} 

\newcommand{\liq}{\ensuremath{L}}
\newcommand{\liqfunction}{\ensuremath{\mathbf{L}}}

\newcommand{\rprice}{\ensuremath{\pi}}
\newcommand{\rpl}{\ensuremath{\pi^l}}

\newcommand{\rpu}{\ensuremath{\pi^u}}

\newcommand{\xreal}{\ensuremath{x^r}}
\newcommand{\yreal}{\ensuremath{y^r}}

\newcommand{\liqtotal}{\ensuremath{\overline{L}}}
\newcommand{\size}{\ensuremath{\alpha}}

\newcommand{\wealth}{\ensuremath{Y}}
\newcommand{\wealthfunction}{\ensuremath{\mathbf{Y}}}
\newcommand{\feesconst}{\ensuremath{c^f}}

\newcommand{\fees}{\ensuremath{\Sigma^f}}
\newcommand{\gas}{\ensuremath{c^g}}

\newcommand{\xhelp}{\ensuremath{\Tilde{x}^r}}
\newcommand{\yhelp}{\ensuremath{\Tilde{y}^r}}

\newcommand{\ctransact}{\ensuremath{c^{tr}}}

\DeclareMathOperator{\EX}{\mathbb{E}}
\newcommand\thefont{\expandafter\string\the\font}

\definecolor{varcolorred}{rgb}{1,0,0}  
\definecolor{varcolorblue}{rgb}{0,0,1}  
\definecolor{varcolorblack}{rgb}{0,0,0}  
\newcommand{\varcolorred}{varcolorred} 
\newcommand{\varcolorblue}{varcolorblue} 

\usepackage{marginnote}
\usepackage{draftwatermark}
\SetWatermarkText{Preprint.\parbox{14.8cm}{\footnotesize{\textcopyright 2025 IEEE. Personal use of this material is permitted. Permission from IEEE must be obtained for all other uses, in any current or future media, including reprinting/republishing this material for advertising or promotional purposes, creating new collective works, for resale or redistribution to servers or lists, or reuse of any copyrighted component of this work in other works.}}}\SetWatermarkColor[gray]{0.5}
\SetWatermarkFontSize{1cm}
\SetWatermarkAngle{0}
\SetWatermarkHorCenter{10.85cm}
\SetWatermarkVerCenter{26.5cm}

\def\BibTeX{{\rm B\kern-.05em{\sc i\kern-.025em b}\kern-.08em
		T\kern-.1667em\lower.7ex\hbox{E}\kern-.125emX}}
\begin{document}

     \title{Automated Market Makers:\\A Stochastic Optimization Approach for Profitable Liquidity Concentration}

\author{
    \IEEEauthorblockN{
        Simon Caspar Zeller\IEEEauthorrefmark{1},
        Paul-Niklas Ken Kandora\IEEEauthorrefmark{1},
        Daniel Kirste\IEEEauthorrefmark{2},
        Niclas Kannengießer\IEEEauthorrefmark{1}
    }
    \IEEEauthorblockN{
        Steffen Rebennack\IEEEauthorrefmark{1},
        Ali Sunyaev\IEEEauthorrefmark{2}
    }
    \IEEEauthorblockA{
        \IEEEauthorrefmark{1}Karlsruhe Institute of Technology, Karlsruhe, Germany\\
        \texttt{\{simon.zeller,paul-niklas.kandora,niclas.kannengiesser,steffen.rebennack\}@kit.edu}\\
    }
    
    \IEEEauthorblockA{
        \IEEEauthorrefmark{2}Technical University of Munich, Campus Heilbronn, Germany\\
        \texttt{\{daniel.kirste, sunyaev\}@tum.de}\\
    }
}

    \maketitle

    \begin{abstract}
Concentrated liquidity automated market makers (AMMs), such as Uniswap v3, enable liquidity providers (LPs) to earn liquidity rewards by depositing tokens into liquidity pools. However, LPs often face significant financial losses driven by poorly selected liquidity provision intervals and high costs associated with frequent liquidity reallocation.
To support LPs in achieving more profitable liquidity concentration, we developed a tractable stochastic optimization problem that can be used to compute optimal liquidity provision intervals for profitable liquidity provision.
The developed problem accounts for the relationships between liquidity rewards, divergence loss, and reallocation costs.
By formalizing optimal liquidity provision as a tractable stochastic optimization problem, we support a better understanding of the relationship between liquidity rewards, divergence loss, and reallocation costs. Moreover, the stochastic optimization problem offers a foundation for more profitable liquidity concentration.
\end{abstract}

    \section{Introduction}
\label{sec:introduction}

In decentralized exchanges (DEXs), automated market makers (AMMs) enable market participants to transparently trade assets represented as digital tokens~\cite{kirste2023tmp}.
Liquidity-concentrating liquidity provider (LP)-based AMMs, such as those used in Uniswap~v3, are a specific type of AMM that sources tokens for trades from liquidity pools~\cite{hayden_uniswap_2021}.
Liquidity pools are software components that store tokens deposited by LPs, such as individuals and organizations.
In return for token deposits, LPs receive \textit{liquidity rewards}---a share of the AMM's market-making revenue~\cite{xu_sok_2023, kirste2023tmp}.
Although gaining profits from liquidity provisioning may seem like a simple income source, most LPs suffer financial losses \cite{milionis_automated_2023, aoyagi_liquidity_2020, cartea2024decentralized}.

For profitable liquidity provisioning, LPs need to understand complex relationships between liquidity rewards, divergence loss, and reallocation costs~\cite{cartea2024decentralized}.
When opening a liquidity position by depositing tokens in a liquidity-concentrating LP-based AMM, LPs must carefully specify the liquidity provision interval for that position.
Liquidity provision intervals are token price ranges in which an AMM can use an LP's deposited tokens for market making~\cite{hayden_uniswap_2020}.
If the token price falls within the provision intervals, the LP's liquidity position is active, and the LP gains liquidity rewards when the AMM uses their tokens for market making. If the token price moves outside the provision intervals, the AMM cannot use the tokens for market making. In this case, the LP's liquidity position becomes inactive, and the LP does not receive liquidity rewards~\cite{hayden_uniswap_2021}.

LPs can increase liquidity rewards by concentrating liquidity in narrow liquidity provision intervals~\cite{hayden_uniswap_2021, echenim:hal-04214315}. 
Narrow liquidity position intervals, however, amplify divergence loss \cite{hashemseresht2022concentrated}. Divergence loss occurs when the value of tokens held by LPs in a liquidity pool diminishes due to price changes compared to the value they would have retained if the tokens were held outside the pool~\cite{milionis_automated_2023}.
Divergence loss decreases if the pool rebalances to the original token ratio when the LP made their deposit~\cite{milionis_automated_2023, xu_sok_2023}.

Narrow liquidity provision intervals not only amplify divergence loss but also increase the risk of token prices moving beyond the specified range when compared to wider intervals~\cite{milionis_automated_2023}.
To provide liquidity with an active position and gain liquidity rewards, LPs can reallocate their liquidity to a new interval around the token price.
Liquidity reallocation, however, incurs reallocation costs, comprised of transaction costs charged by the blockchain system and rebalancing costs in the form of trading costs charged by the AMM~\cite{cartea2024decentralized}.

To help LPs find optimal liquidity provision intervals to increase profits despite divergence loss and reallocation costs, extant research offers optimization approaches for profitable liquidity provision based on token volatility and token prices\cite{cartea2024decentralized, fan_differential_2022, fan2021strategic}.
These approaches, however, rely on continuous-time models with continuous reallocation, which deviates from practical implementations due to the discrete nature of real-world reallocation, neglect reallocation, or treat the interval width as fixed rather than optimizing it.

To better account for reallocation costs when computing optimal liquidity provision intervals, we approach the following research question:
\textit{What is a useful approach to support LPs in finding optimal liquidity provision intervals to increase profitability?}

We first formalized liquidity concentration and the readjustment of liquidity provision intervals as a tractable stochastic optimization problem. 
Second, we used the optimization problem to devise an optimal liquidity reallocation strategy and applied the strategy to historical on-chain trading data of Uniswap~v3.
Comparing the devised strategy with other strategies showcases the utility of the tractable stochastic optimization problem to optimize liquidity provision intervals and their readjustments. 

Our main ambition in this work is to support LPs in developing more profitable liquidity concentration strategies.
By presenting a tractable stochastic optimization problem that optimizes the width of liquidity provision intervals and incorporates its influence on liquidity rewards, divergence loss, and reallocation costs, we support a better understanding of the trade-off between liquidity rewards, divergence loss, and reallocation costs.
This contributes to increasing the profitability of liquidity provision and the efficiency of liquidity-concentrating AMMs.

The remainder of this work is structured into five sections. Section~\ref{sec:background} explains the foundations of liquidity-concentrating AMMs relevant to this work and describes related research.
Next, section~\ref{sec:optimization} presents a formalization of profitable liquidity concentration in the form of an optimization problem. 
The utility of the formalization is then demonstrated in section~\ref{sec:demo}.
In section~\ref{sec:discussion}, we discuss the principal findings and explain the main contributions of this work and its limitations. Moreover, we outline future research directions.
Section~\ref{sec:conclusion} concludes this work with a brief summary and key takeaways from this work.

    \section{Background}
\label{sec:background}

This section first elucidates the foundations of liquidity-concentrating LP-based AMMs that are important in this work.
Next, foundations of divergence loss and reallocation costs in liquidity concentration are explained.

\subsection{Automated Market Makers}
\label{sec:cryptoeconomic-systems}

AMMs are software-based market makers that continuously quote bid (buy) and ask (sell) prices to exchange tokens with market participants in cryptoeconomic systems~\cite{kirste2023tmp, xu_sok_2023}.
To compute token prices, AMMs use mathematical functions implemented in software code of smart contracts \cite{kirste2023tmp}. Because smart contract code is visible to users with access to a blockchain system \cite{kannengiesser2021smartcontract}, pricing functions are transparent.
AMMs use price discovery mechanisms to compute token exchange rates.
This work focuses on AMMs with constant-product price discovery as a special type of AMM. These AMMs ensure that the product of the token quantities in a liquidity pool is invariant during trades \cite{hayden_uniswap_2020, xu_sok_2023}.

To provide liquidity to markets, AMMs require market participants to act as LPs by depositing tokens in liquidity pools~\cite{kirste_undergirding_2025}.
To motivate market participants to deposit tokens, AMMs distribute fractions of collected market-making fees to LPs in the form of liquidity rewards.
The amount of liquidity rewards is determined by the market-making fees collected and the proportion of an LP's liquidity relative to the total liquidity available to the AMM~\cite{xu_sok_2023}.

Basic AMMs like Uniswap~v2 distribute provided liquidity uniformly across all possible token prices~\cite{hayden_uniswap_2020}.
This uniform distribution ensures that the same amount of liquidity is available for every possible token price. However, liquidity is primarily needed within a narrow range around the current token price. As a result, liquidity positioned far from this range remains largely unused, making uniform distribution inefficient~\cite{xu_sok_2023}.

To more efficiently provide liquidity, more advanced constant-product AMMs (e.g.,~Uniswap~v3) enable LPs to concentrate provided liquidity by defining intervals for positions of deposited tokens~\cite{hayden_uniswap_2021, loesch_impermanent_2021}.
Liquidity concentration enables more efficient liquidity allocation, increasing the pool’s liquidity within the LP’s chosen intervals~\cite{hayden_uniswap_2021, echenim:hal-04214315}.
AMMs only use tokens from active positions for market making. A position is active when the token price is within the liquidity provision intervals~\cite{hayden_uniswap_2021}.

Liquidity concentration also benefits LPs. LPs can deposit tokens in narrow price intervals to amplify their share of liquidity rewards by increasing their share of provided liquidity within those intervals~\cite{cartea2024decentralized, milionis_automated_2023}.
While offering the chance to gain more rewards, narrow liquidity provision intervals increase  the risks of divergence loss and opportunity costs, as inactive liquidity misses out on trading opportunities when prices move beyond the defined range~\cite{cartea2024decentralized}.
Profitable liquidity provisioning thus requires careful selection and readjustment of liquidity provision intervals~\cite{cartea2024decentralized}.

The adjustment of liquidity provision intervals entails reallocation costs, which comprise costs charged by the blockchain system for transaction processing and rebalancing costs charged by the AMM~\cite{cartea2024decentralized}. Such costs can erode profits of LPs, especially those of LPs that provide lower amounts of tokens.
For profitable liquidity provisioning, LPs must actively monitor market conditions and readjust their liquidity provision intervals to match token price developments.

\subsection{Divergence Loss and Reallocation Costs in AMMs}
\label{sec:readjustment_cost}

Divergence loss occurs when token prices in a liquidity pool change after a deposit. For example, when a token $A$ increases in price compared to a token $B$ in the same liquidity pool, the token holdings in the liquidity pool change. This happens because the price reflects the ratio of tokens $A$ and $B$.
After this process, LPs own more of the lower-valued token $B$ and less of the higher-valued token $A$.
LPs withdrawing under these conditions suffer a financial loss compared to holding the deposited tokens outside of the liquidity pool.
Divergence loss decreases if the liquidity pool recovers to the original token ratio from when the LP made their deposit~\cite{milionis_automated_2023, xu_sok_2023}. Narrower liquidity provision intervals are prone to higher divergence loss \cite{hashemseresht2022concentrated}.

Compared to wider intervals, narrow liquidity provision intervals typically require more frequent liquidity reallocation to gain liquidity rewards~\cite{fan2021strategic}.
Reallocation incurs reallocation costs that comprise \textit{transaction costs} and \textit{rebalancing costs}. 
\textit{Transaction costs} arise from fees that a blockchain system charges for transaction processing.
\textit{Rebalancing costs} occur when LPs rebalance their token holdings for liquidity reallocation.
In liquidity reallocation, LPs must first close an existing position and withdraw their tokens. Then, to open a new position, LPs must hold the tokens in a specific ratio, which often differs from the ratio of their withdrawn tokens.
For example, if the price exceeds the upper limit of a liquidity provision interval, the LP will possess only token $B$. To open a new position with liquidity provision intervals around the token price, the LP must provide a specified quantity of both tokens $A$ and $B$. Consequently, the LP needs to exchange tokens $B$ for token $A$, incurring fees charged by the AMM.
To own assets in that ratio, LPs rebalance their token holdings by exchanging tokens through the AMM's liquidity pool. However, using the AMM's liquidity pool incurs trading fees charged by the AMM. The fees are proportional to the amount of assets that the LP exchanges~\cite{hayden_uniswap_2021}.

For profitable liquidity provisioning, LPs must find a Pareto-optimum regarding the trade-off between liquidity rewards, divergence losses, and reallocation costs~\cite{Heimbach2022, kirste_influence_2024}. Narrow intervals increase liquidity rewards but increase the likelihood of the token price falling outside the interval, requiring more frequent reallocations and, therefore, higher reallocation costs. Token price movements additionally lead to divergence losses, which are amplified by narrower liquidity provision intervals and reduce overall returns for LPs. Conversely, broader intervals decrease liquidity rewards but also decrease divergence loss and the likelihood of the token price moving out of the liquidity provision intervals, thus decreasing reallocation costs.
LPs must, therefore, appropriately balance the trade-off between liquidity rewards, divergence loss, and reallocation costs by selecting an optimal interval width.

\subsection{Related Work on Optimal Liquidity Provision Strategies}

Extant research has made valuable contributions to developing profitable liquidity provision strategies.
Cartea et al. (2024) propose a mathematical model for optimal liquidity provisioning that can be used to compute optimal liquidity provision intervals based on price volatility, price drift, and pool profitability~\cite{cartea2024decentralized}. They present closed-form solutions based on a continuous-time model of liquidity provision with continuous reallocation. In practice, however, reallocation takes place at discrete points in time.

Fan et al. (2022) present profit and loss models for Uniswap~v2 and v3, which solve sample average approximation (SAA)-based optimization problems but neglect liquidity reallocation~\cite{fan_differential_2022}. Additionally, Fan et al. (2023) propose an approach to compute optimal liquidity allocation by maximizing SAA-approximated terminal wealth. They compute the optimal allocation of liquidity for a fixed reallocation rule that only depends on price movements. In their optimization approach, however, they do not optimize the interval width. Instead, they optimize the allocation of liquidity within a given interval of fixed size and, therefore, a fixed expected reallocation frequency~\cite{fan2021strategic}.
    \section{Profitable Liquidity Reallocation as an Optimization Problem}
\label{sec:optimization}

This section first presents a mathematical description of liquidity-concentrating constant-product AMMs and reallocation costs. We use the mathematical description of constant-product AMMs and reallocation costs to represent the computation of liquidity provision intervals as a stochastic optimization problem, which we solve with commercial software. 
The optimization problem and its solution are envisioned to support LPs in developing more profitable liquidity provision strategies.

\label{sec:optimization-model}
\subsection{Mathematical Description of Constant Product Market Makers}

Liquidity pools of constant-product AMMs represent reserves of token pairs $\left(x,\,y\right)$. The total liquidity of a pool is denoted as $\liqtotal$ and calculated as $\liqtotal \coloneqq \sqrt{x y}$. During token exchanges, the liquidity $\liqtotal$ of a pool is constant. An AMM exchanges $\Delta^x$ units of $x$ for $\Delta^y$ units of $y$ so that $\liqtotal$ remains constant: $\liqtotal = \left(x + \Delta^x,\, y - \Delta^y\right)$~\cite{hayden_uniswap_2021}.

The marginal exchange rate $y \mathbin{/} x$ represents the price of one token $x$ in terms of the other token $y$~\cite{hashemseresht2022concentrated}. We will simply refer to this as the price and use square root prices $\rprice$ to simplify further notation.

Uniswap~v3~\cite{hayden_uniswap_2021} is a special type of constant-product AMM that enables LPs to concentrate liquidity. In Uniswap~v3, each LP must define liquidity provision intervals $[\rpl,\rpu]$ to open a liquidity position.
If the token price falls within the interval of a position, the position is active, and the corresponding LP earns liquidity rewards proportional to its share of the total liquidity available in the pool for the token price~\cite{hayden_uniswap_2021}.

Liquidity positions can be expressed through a triple $(\liq, \rpl, \rpu)$, where $\liq$ quantifies liquidity within the position's liquidity provision interval. Within $[\rpl,\rpu]$, deposited tokens are treated as \textit{virtual reserves} to amplify the LP's provided liquidity in the liquidity provision interval. Uniswap~v3 distinguishes between \textit{real reserves} and \textit{virtual reserves}.
Real reserves refer to token amounts held by an LP, which we will denote in the following as $\left(\xreal,\, \yreal\right)$.
Virtual reserves represent liquidity concentrated within $[\rpl, \rpu]$.
Real reserves associated with a liquidity position and square root price $\rprice$ can be computed with the following expression~\cite{Heimbach2022}:

\begin{equation}
\label{eq:realassets}
    \left(\xreal,\, \yreal\right) =
    \begin{cases}
        \left(\liq  \left(\frac{1}{\rpl}-\frac{1}{\rpu}\right),\, 0\right) & \text{if } \rprice < \rpl, \\
        \left(\liq  \left(\frac{1}{\rprice}-\frac{1}{\rpu}\right),\, \liq \left(\rprice-\rpl\right)\right) & \text{if } \rpl \leq \rprice \leq \rpl \\
        \left(0,\, \liq \left(\rpu-\rpl\right)\right) & \text{if } \rprice > \rpu.
    \end{cases}
\end{equation}

For $\left(\xreal,\, \yreal\right)$, \cref{eq:realassets} helps LPs figure out how much liquidity $\liq$ they can provide based on $\rprice$ and a liquidity provision interval $[\rpl, \rpu]$. Before providing liquidity, the LP adjusts token amounts to be provided to ensure they are in the ratio satisfying \cref{eq:realassets}.

\begin{table}[!b]
    \centering
    \begin{tabular}{p{0.12\linewidth} | p{0.78\linewidth}}
        \toprule
        \textbf{Symbol} & \textbf{Description} \\
        \midrule
        $x$         & First token \\
        $y$         & Second token \\
        $\rprice_t$ & Square root of the marginal exchange rate \\
        $\liq_t$    & Liquidity provided by the LP \\
        $\size$     & Relative size of the square root price interval \\
        $\gamma$    & Threshold that triggers reallocation \\
        $\rpl_t$    & Lower bound of the LP positions square root price interval \\
        $\rpu_t$    & Upper bound of the LP positions square root price interval\\
        $\xreal_t$  & Real tokens $x$ held by the LP \\ 
        $\yreal_t$  & Real tokens $y$ held by the LP \\
        $\wealth_t$ & Wealth of the LP \\
        $\fees_t$   & Accumulated and not-yet-claimed liquidity rewards \\
        $\gas$    & Transaction fees paid per interaction \\
        $\ctransact$ & Fees paid by traders per unit of trade volume \\
        $\overline{\feesconst}_t$ & Liquidity rewards paid to LP per time step and unit of provided liquidity, if their position is permanently active \\
        $\feesconst_t$ & Liquidity rewards paid to LP per time step and unit of provided liquidity \\
        $L_t$ & Liquidity provided by an LP \\
        $\overline{L}_t$ & Liquidity of the pool \\
        $T$ & Horizon for liquidity provision \\
        \bottomrule
    \end{tabular}
    \vspace{1pt}
    \caption{List of important Symbols}
    \label{tab:variable}
\end{table}

\subsection{Readjustment and Cost Implications}

LPs earn liquidity rewards proportional to their share of the pool's total liquidity, represented as $\liq \mathbin{/} \liqtotal$. To increase liquidity rewards, LPs concentrate their liquidity within narrow price ranges $[\rpl,\rpu]$, which increases $\liq$ within these ranges.
When the current $\rprice$ moves outside $[\rpl,\rpu]$, however, the liquidity position becomes inactive.
LPs must reallocate their liquidity to earn liquidity rewards, which incurs reallocation costs (see \cref{sec:readjustment_cost}). We denote transaction costs per reallocation as $\gas$ and rebalancing costs per amount of exchanged tokens as $\ctransact$.

\subsection{Liquidity Provision as an Optimization Problem}
\label{sec:LPasoptimization}

We focus on a single LP to model the problem of optimal reallocation while considering liquidity rewards, divergence loss, and reallocation costs.
The LP starts with initial wealth $Y_0$ given by the value of token holdings $\xreal_0$ and $\yreal_0$. We assume their token holdings to be in the correct ratio for liquidity provision, i.e. $\yreal_0=\rprice_0^2 \xreal_0$. The LP then uses all their wealth to provide liquidity from $t=0$ to $t=T$. Given $\rpl_0$ and $\rpu_0$, the initial tokens can be translated into initial liquidity $\liq_0$ using \cref{eq:realassets} (and vice versa).

Similar to \cite{cartea2024decentralized}, we model the marginal price $\rprice_t^2$ as a Geometric Brownian motion (GBM) process with constant volatility $\sigma$ but without drift because forecasting price trends is beyond the scope of this work. 
The optimal provision interval is expressed as $\left[\frac{1}{\size}\rprice, \size\rprice\right]$ for an optimal $\alpha>1$ which is Pareto-optimal regarding the trade-off between liquidity rewards, divergence loss and reallocation costs \cite{cartea2024decentralized}. In this work, $\alpha$ is independent of time.

In line with existing research \cite{fan2021strategic}, we assume the LP to reallocate their liquidity at time $t$ when the liquidity position becomes inactive: $\rprice_t \notin \left[\rpl_{t-1},\, \rpu_{t-1}\right]$
As long as $\rprice_t \in \left[\rpl_{t-1},\, \rpu_{t-1}\right]$, the liquidity position remains unchanged to avoid reallocation costs.
The proposed strategy, therefore, can be interpreted as an adapted version of a strategy that \cite{fan2021strategic} refer to as \textit{ULRA}.

To compute liquidity rewards, we calculate the proportion of time between $t-1$ and $t$ during which the position is active, meaning $\rprice \in [\rpl_{t-1}, \rpu_{t-1}]$. The calculated proportion is based on linearly interpolated prices and is denoted as $p_t$.
The proportion $p_t$ is multiplied by $\overline{\feesconst}_t$, which describes the liquidity rewards collected per unit of liquidity between $t-1$ and $t$ if the position were continuously active.
We therefore express the liquidity rewards earned by the LP between $t-1$ and $t$ as $p_t \overline{\feesconst}_t \liq_{t-1} = \feesconst_t \liq_{t-1}$.
The LP does not change their active liquidity positions. The liquidity rewards are therefore collected until the LP reallocates their liquidity. This involves closing the existing position and reopening a new position using the accumulated liquidity rewards and the withdrawn tokens.

Terminal wealth (i.e., the value of token holdings together with the collected and not-yet-claimed liquidity rewards measured in USD) is maximized by determining the optimal liquidity provision interval size $\alpha$.
We use $z_t$ as a binary variable that indicates whether or not the LP reallocates at time $t$. Due to the continuity of $\rprice_t$, it does not matter whether we reallocate if $\rprice_t \notin \left[\rpl_{t-1},\, \rpu_{t-1}\right]$ or $\rprice_t \notin \left(\rpl_{t-1},\, \rpu_{t-1}\right)$. However, we use the second reallocation rule as it is easier to implement in an optimization problem. 
This leads to the following optimization problem:

\begin{subequations} \label{optproblem}
{\allowdisplaybreaks
\begin{align} 
    &\textbf{P}: \max_{\substack{\alpha}} 
     \EX\left[\wealthfunction_T\left(\rprice_T, \liq_{T-1}, \rpl_{T-1},  \rpu_{T-1}, \fees_T \right) \right] \label{optproblem_obj}  \\[5pt]
    &\text{s.t.} \quad \rprice_t \geq \rpu_{t-1} \text{ or } \rprice_t \leq \rpl_{t-1} \Rightarrow z_t = 1 \text{ else } 0 , \label{optproblemz_t} \\ 
    & \rpl_t = z_t \cdot \frac{1}{\alpha}\rprice_t + \left(1-z_t\right) \cdot \rpl_{t-1}  ,  \label{optproblemrpl} \\ 
    & \rpu_t = z_t \cdot \alpha\rprice_t + \left(1-z_t\right) \cdot \rpu_{t-1}  ,   \label{optproblemrpu} \\ 
    & \fees_t = \fees_{t-1} \left(1-z_{t-1}\right) + \feesconst_t \liq_{t-1},  \label{optproblemfees} \\
    & \liq_t = \liqfunction_t\left(\liq_{t-1},  \rpl_{t-1},  \rpu_{t-1},  \rprice_t,   \rpl_t,  \rpu_t,  \fees_{t},  \ctransact,  \gas \right), \label{optproblemliq} \\  
    & \alpha \geq 1 + \epsilon, \label{optproblemalpha} 
\end{align}}
\end{subequations}

$\wealthfunction_t$ and $\liqfunction_t$ refer to functions that map the parameters to $\wealth_t$ and $\liq_t$ (see \cref{eq:realassets}). Equations~\eqref{optproblemz_t}-\eqref{optproblemrpu} and \eqref{optproblemliq} must hold for $t=1,\, \dots ,\,T-1$ and \cref{optproblemfees} for $t=1,\, \dots ,\,T$. 
We introduce $\epsilon>0$ together with~\eqref{optproblemalpha} for numerical reasons and because it is not feasible to choose an arbitrarily small liquidity provision interval width within Uniswap~v3 \cite{hayden_uniswap_2021}.
Formulation~\eqref{optproblem_obj}--~\eqref{optproblemalpha} constitutes a mixed-integer non-linear problem which can be solved using commercial software, such as Gurobi 12~\cite{gurobi2023}. For additional information concerning the tractable problem formulation of equations~\eqref{optproblem_obj}--~\eqref{optproblemalpha} and implementation details, we refer the reader to \cref{optproblemgurobi} in~\cref{sec:gurobiformulation}.

    \section{Utility Demonstration}
\label{sec:demo}
To illustrate the utility of the optimization problem for optimal liquidity provision, we analyzed on-chain data from Uniswap~v3. We first analyze the profitability of LPs providing liquidity as described in \cref{sec:LPasoptimization} for different values of $\size$. Next, we compute an optimal solution to the stochastic optimization problem using commercial software and validate the solution.
In \cref{sec:methods}, we describe the approach employed for the demonstration. The results are presented in \cref{sec:results}.

\subsection{Methods}
\label{sec:methods}
\begin{figure}[ht] 
\centering
\includegraphics[scale=1.0]{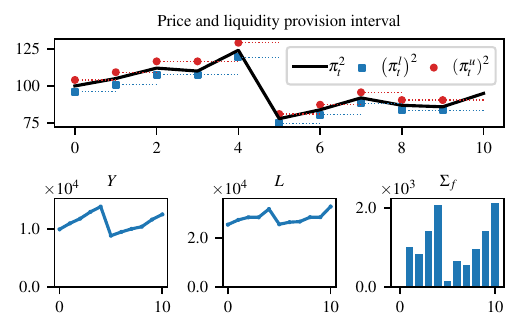} 
\caption{Illustration of the evolution of an LP's liquidity provision interval, wealth, liquidity and unclaimed fees.}\label{fig:illustrationLPposition}
\end{figure}

\begin{figure}[ht] 
\centering
\includegraphics[scale=1.0]{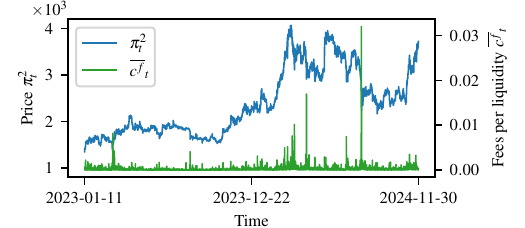} 
\caption{Price and liquidity rewards per unit of virtual liquidity ($L$) during the analyzed time period.}\label{fig:priceandfees}
\end{figure}

Following the modeling assumptions outlined in \cref{sec:LPasoptimization}, we consider LPs with an initial token wealth $Y_0 = \rprice_0^2\xreal_0+\yreal_0$, providing liquidity in the manner described in the same section. \Cref{fig:illustrationLPposition} illustrates the strategy. To further analyze the impact of the event triggering reallocation, we also formulate a modified strategy. In this strategy, liquidity is reallocated when $\rprice_t < \rpu_{t-1} - \gamma$ or $\rprice_t > \rpu_{t-1} + \gamma$. The LP therefore reallocates liquidity sooner when $\gamma < 0$, whereas it delays reallocation when $\gamma > 0$. 
For the demonstration, we used the $0.05\%$ USDC/ETH liquidity pool on Uniswap~v3---one of the most adopted pools by trade volume~\cite{uniswap_uniswap_2024}. We consequently set $\ctransact$ to 0.05\%. Our dataset spans from January 11, 2023, starting with the Ether price at approximately \$1,350, to November 30, 2024, with a price of \$3,725. The dataset is compiled from hourly data and contains the price together with \textit{feeGrowthGlobal0X128} ($f_{g, 0}$) and \textit{feeGrowthGlobal1X128} ($f_{g, 1}$).
The variables $f_{g,0}$ and $f_{g,1}$ track the cumulative liquidity rewards that LPs would have earned from a position with liquidity $1$ that has been active since the inception of the AMM \cite{hayden_uniswap_2021}. Estimates for $\overline{\feesconst}_t$ are derived by calculating the changes in $f_{g, 0}$ and $f_{g, 1}$ between times $t-1$ and $t$ and converting them into USDC. We consequently assume the LP to neither influence the price development nor the liquidity rewards earned per unit of liquidity ($f_{g, 0}$ and $f_{g, 1}$). Both variables are illustrated in \cref{fig:priceandfees}.

Similar to \cite{fan_differential_2022}, we use SAA \cite{kleywegt2002sample} to estimate the objective function in \cref{optproblem}. This involves generating random price paths by simulating GBM, which are then handed to the formulation of the optimization problem described in the appendix (\hspace{1sp}\cref{optproblemgurobi}).
As we used SAA to solve the optimization problem, the simulation is stochastic and not deterministic. To obtain reliable results, we solved the problem multiple times for different random seeds.

The standard deviation of the GBM was estimated using the standard deviation of the log return of the price data and the drift was set to zero. Transaction costs were estimated to be \$ 109.8 per reallocation (\$ 84.8 as of \cite{cartea2024decentralized} plus an additional estimated \$ 25 to claim liquidity rewards). The maximum liquidity rewards per liquidity $\overline{c^f}$ are set to the median of $\overline{c^f}_t$, and consequently assumed to be constant. The median was chosen as a robust estimator, as the data shows very large values of $\overline{\feesconst}_t$ at \textit{price jumps}.

We bounded $\size$ to avoid numerical issues and to efficiently solve the optimization problem. The lower bound was set to $1.01$  because values for $\size$ below $\approx 1.02$ consistently resulted in profits of $-100\%$ in the used data set. We set the upper bound to $4$, as values of $\size$ above $1.73$ result in no reallocation taking place over the entire time frame of the used data.
Additionally, already for $\size=4$, the associated price interval covers movements ranging from $1 \mathbin{/} 16$ to $16$ times the initial price. The probability that any of the sampled price paths leaves the initial bounds is, therefore, very close to zero, given that the hourly $\sigma$ of the GBM was estimated to $\approx 0.006$ and paths are simulated for 5-10 steps. Reallocation costs therefore do not influence terminal wealth and the solution $\size=4$ can be interpreted as going full range.

Given an initial wealth of the LP, the optimization problem is solved using a formulation of \cref{optproblemgurobi} with SAA \cite{kleywegt2002sample}, that we solve with Gurobi \cite{gurobi2023}. Both logical constraints and the linearization of binary-continuous variable products are implemented using Big-M constraints \cite{williams2013model}. The code can be accessed at \url{https://github.com/simonczeller/AMM_Liquidity_Concentration_Optimization}.

\subsection{Demonstration Results}
\label{sec:results}

\begin{figure}[ht] 
\centering
\includegraphics[scale = 1.0]{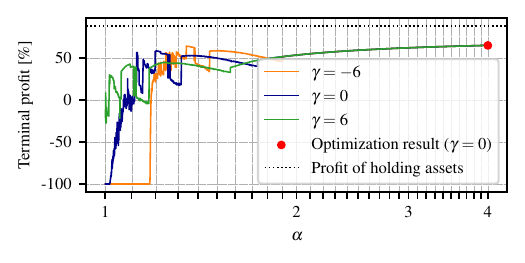}
\caption{Terminal profit of an LP with initial wealth \$ 100,000 for different sizes $\size$ of the provision interval and values of $\gamma$, expressed as a percentage of the initial wealth.}\label{fig:profitsfordifferentalphaandgamma}
\end{figure}

\Cref{fig:profitsfordifferentalphaandgamma} shows the profits an LP with an initial wealth of \$ 100,000 would have made if they provided their entire wealth in the Uniswap~v3 0.05\% pool for different liquidity provision interval sizes $\alpha$ from January 11, 2023 to November 30, 2024.
\begin{figure}[ht] 
\centering
\includegraphics[scale = 1.0]{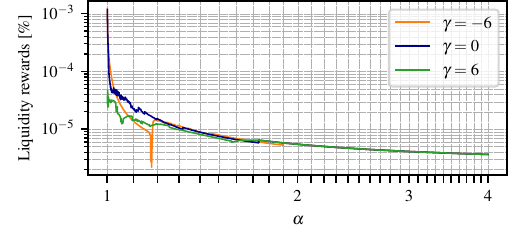}
\caption{Liquidity rewards per time step of an LP with initial wealth \$ 100,000 for different sizes $\size$ of the provision interval and values of $\gamma$, expressed as a percentage of the initial wealth.}\label{fig:feesfordifferentalphaandgamma}
\end{figure}
\begin{figure}[ht] 
\centering
\includegraphics[scale = 1.0]{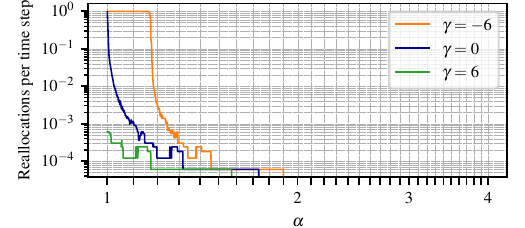}
\caption{Profits of an LP with initial wealth \$ 100,000 for different sizes $\size$ of the provision interval and values of $\gamma$.}\label{fig:reallocationsfordifferentalphaandgamma}
\end{figure}
The profits are shown for various liquidity provision interval sizes ranging from $\size=1.0001$ to $\alpha=4$. The corresponding average liquidity rewards and average reallocations per time step are shown in \cref{fig:feesfordifferentalphaandgamma} and \cref{fig:reallocationsfordifferentalphaandgamma}. 
Liquidity rewards (or reallocations) per time step are calculated as follows: If profits are greater than $-100\%$, the total liquidity rewards (or reallocations) are divided by the total number of time steps in the analyzed data. If profits reach $-100\%$, the denominator is given by the number of time steps until the LP has no remaining wealth.
We will first discuss the results for $\gamma=0$, to then examine the influence of a positive and negative value of $\gamma$.
\Cref{fig:profitsfordifferentalphaandgamma} shows that for very small liquidity provision interval sizes $\size$, despite very large liquidity rewards (\hspace{1sp}\cref{fig:feesfordifferentalphaandgamma}), the increased divergence loss together with the frequency of reallocations (\hspace{1sp}\cref{fig:reallocationsfordifferentalphaandgamma}) with associated reallocation costs ultimately result in negative profits. For an increased $\size$ greater than $\approx 1.11$, liquidity provision would have been profitable. Returns through liquidity returns and price increases therefore would have offset reallocation costs.  

The jump in \cref{fig:profitsfordifferentalphaandgamma} at $\size \approx 1.3$ is explained by only one reallocation taking place over the entire time-frame for $\size>1.3$. This results in higher profits through decreased costs associated with reallocation.
For $\size \geq 1.73$, the lower bound $\left(\rpl_0\right)^2$ is smaller than the minimum price and the upper bound $\left(\rpu_0\right)^2$ is greater than the maximum price. Thus, the initial liquidity provision interval is large enough such that no reallocation is needed (see \cref{fig:reallocationsfordifferentalphaandgamma}). The steady increase of profits for $\alpha \geq 1.73$ can be explained by the reduction of the divergence loss for larger liquidity position sizes~\cite{hashemseresht2022concentrated}. In this case, this effect outweighs lower liquidity rewards due to a decrease in virtual liquidity. Going full range would have resulted in profits of $71\%$, compared to $65\%$ for $\size=4$.

Liquidity provisioning is profitable in the analyzed period, mainly because the price of ETH nearly tripled. Conversely, a decrease in the price of ETH could have reduced profitability, potentially leading to financial losses depending on the magnitude of the price decrease and the specific path. A decrease (compared to an increase) would, however, not influence the number of reallocations, as they are mainly driven by volatility. Likewise, it would not affect the influence of $\size$ on DL which predominantly describes the profitability of larger positions.
Consequently, the significant increase in the price of ETH in the analyzed period does not substantially bias the results.

Liquidity rewards were insufficient to offset LPs' divergence loss.
The dotted line in \cref{fig:profitsfordifferentalphaandgamma} illustrates the profits an LP would have gained by holding the original tokens--half of his wealth in USDC and half in ETH--rather than providing liquidity.

We solved the optimization problem (see \cref{optproblem}) for an initial wealth of \$ 100,000 and different combinations of $T$ and number of sampled price paths $S$. Because the solution to the optimization problem is not deterministic as we use SAA, we have solved the problem multiple times to obtain reliable results. The values in \Cref{tab:descriptive_statistics_alpha} show that the optimal solution was consistently set to $\approx 4$; equivalent to the value for $\alpha$ that an LP should have chosen in retrospect. 

For $\gamma=0$, the LP reallocates whenever $\rprice_t \notin \left[\rpl_{t-1},\, \rpu_{t-1} \right]$, despite a chance of the price reentering the interval. For this reason, it may be beneficial to reallocate only if $\rprice_t$ exceeds a specific distance to the interval.
However, reallocating earlier reduces the probability of $\rprice_{t+1} \notin \left[\rpl_{t},\, \rpu_{t} \right]$ and therefore the expected time the position remains active, potentially increasing liquidity rewards.
To evaluate such effects, the simulation was conducted for $\gamma=6$ and $\gamma=-6$ and reallocation to be triggered if $\rprice_t < \rpu_{t-1} - \gamma$ or $\rprice_t > \rpu_{t-1} + \gamma$. For a negative value of $\gamma$, an increase of liquidity rewards is observed for very small $\size$. However, the increase of reallocations with associated costs ultimately leads to less profitability for smaller values of $\size$. A positive $\gamma$ significantly decreased the number of reallocations, leading to a great increase in profitability for smaller values of $\size$, despite less liquidity rewards.

There is insufficient data to determine whether varying values of $\gamma$ affect profitability when the corresponding $\size$ is optimally chosen.
The data, however, indicates that at least for small values of $\alpha$ a positive $\gamma$ can increase the profitability of liquidity provision.
Additionally, the optimization problem presented in \cref{optproblem} and \cref{optproblemgurobi} can be easily adapted to simultaneously optimize $\size$ and $\gamma$. A simultaneous optimization can be achieved by implementing $\gamma$ as a decision variable and adapting the corresponding constraints. 
    \section{Discussion}
\label{sec:discussion}
This section presents the key findings and contributions of our research, examines major limitations, and explores potential strategies to address these limitations in future studies.

\subsection{Principal Findings}
\label{sec:principal-findings}

The demonstration suggests that providing liquidity within wide liquidity provision intervals is profitable due to the significant increase in the marginal exchange rate.
Due to the low virtual liquidity associated with such interval widths, liquidity rewards are too low to offset divergence loss.
Providing liquidity in narrow intervals yields significantly higher liquidity rewards than wider intervals. Despite increased rewards, these strategies tend to generate lower profits or even lead to negative returns compared to using wider intervals. The reason is an increased divergence loss and more frequent reallocation, leading to higher reallocation costs.
The data shows that LPs improve the profitability of strategies that use narrow liquidity provision intervals by delaying reallocation. LPs achieve this by only reallocating when the price distance to the interval exceeds a specified threshold, denoted as $\gamma$. This approach minimizes unnecessary reallocation costs while maintaining efficient liquidity management.
Overall, even for an optimal $\size$, the demonstration shows that liquidity provision would have been less profitable than not providing liquidity.

Moreover, the demonstration indicates that full-range liquidity provision intervals would have yielded the highest profitability.
We attribute this outcome to the reduced divergence loss associated with wider intervals, combined with the price change observed during the analyzed time frame.

The demonstration shows that the formulation of the tractable stochastic optimization problem helps identify optimal provision intervals. The optimization problem takes the influence of the liquidity provision interval on liquidity rewards, divergence loss and reallocation costs into account and can be solved within a reasonable time using commercial software (see \Cref{tab:descriptive_statistics_runtime}).
The computed optimal liquidity provision interval aligns with the interval an LP should have selected in hindsight. This highlights the model's potential to increase profits by optimizing liquidity rewards while accounting for divergence loss and reallocation costs.

\subsection{Contributions}
\label{sec:contributions}
Our main ambition is to support LPs in developing more profitable liquidity provision strategies, including identifying and readjusting optimal liquidity provision intervals. 
We first formulate a liquidity provision strategy that LPs can implement straightforwardly. Based on this strategy, a tractable stochastic optimization problem is formalized that seeks to optimize liquidity provision interval width. The optimization problem incorporates the three main components that determine the profitability of LPs: liquidity rewards, divergence loss, and reallocation costs. The formalization supports a better understanding of the relationships between the main components of liquidity provision profitability.

The stochastic optimization problem can be solved using commercial software. Moreover, it can easily be adapted to reflect various stochastic models of price movements and other market parameters, as well as user-defined thresholds for reallocation. This flexibility allows LPs to tailor the strategy to specific market conditions and provides a cornerstone for more profitable liquidity provision.

\subsection{Limitations}
\label{sec:limitations}

To formulate liquidity provisioning as an optimization problem, we made multiple assumptions that limit the generalizability of our results.
We assumed that LPs do not influence price movements and the total liquidity in the pool. While the influences on either of these may be negligible for small liquidity positions and larger pools, LPs with large liquidity positions providing liquidity in smaller pools may substantially influence total liquidity and token prices when rebalancing their token holdings.

We analyzed a finite-horizon problem with a limited number of time-steps $T$ due to runtime constraints. However, the resulting optimal value for $\size$ may differ from the truly optimal value for an (infinite-horizon) problem or the maximization of wealth over more than $T$ time steps.

Using the mathematical model, we calculated liquidity rewards by interpolating token prices. This implies that if $\rprice_{t} \in \left[\rpl_{t-1},\, \rpu_{t-1} \right]$, we assume constant liquidity rewards per unit of active liquidity between $t-1$ and $t$, i.e., $\feesconst_t = \overline{\feesconst}_t$. However, there is a nonzero probability that $\rprice_s \notin \left[\rpl_{t-1},\, \rpu_{t-1} \right]$ for some $s \in \left(t-1,\, t\right)$. This can lead to an overestimation of expected liquidity rewards, especially for very small values of $\size$. An alternative to estimate the expected time the position is active between $t-1$ and $t$ is to use Brownian bridges.

We assumed that the token price follows GBM without drift. Still, the optimization problem can easily be adapted to other stochastic processes, including GBM with drift and dynamic volatility, given that $S$ price paths are simulated and handed to the optimization problem as parameters. Such an implementation may require adapting the liquidity provision interval, possibly by introducing time-dependent $\size_t$ or decision variables for the upper and lower bound of the liquidity provision interval.

\subsection{Future Research}
\label{sec:future-research}

We tested the liquidity provision strategy based on on-chain data of Uniswap~v3 based on discrete one-hour time steps.
Additional empirical investigations are necessary to understand better how profitable the proposed strategy is when providing liquidity to other pools with different costs, rewards, or price developments.
Furthermore, the optimization approach presented in this work can be compared with other approaches (e.g.,\cite{cartea2024decentralized}, \cite{fan2021strategic}) on real-world data.

Many algorithms, such as Nested Benders Decomposition or Stochastic Dual Dynamic Programming~\cite{füllner_sddp}, can reduce the runtime of stochastic optimization problems, enabling the solution of large problems in acceptable time. Future research may adapt and implement the formulation of the optimization problem presented in this work for such algorithms to solve the problem for a larger sample size and more time steps.

The optimization approach relies on the accuracy of estimations of parameter values, including liquidity rewards per unit of liquidity, price volatility, and transaction fees. As forecasting those parameter values was beyond the scope of this work, accurate forecasting remains a potential area for future research.

\section{Conclusion}
\label{sec:conclusion}
Liquidity concentration is complex because LPs must find Pareto-optima between liquidity rewards, divergence loss, and reallocation costs to increase profits from liquidity provision.
This work presents a tractable stochastic optimization problem to assist LPs in computing optimal strategies for more profitable liquidity concentration---a foundational activity to the viability of liquidity-concentrating AMMs. The formulation of this problem can help LPs in finding the Pareto-optima between liquidity rewards, divergence loss, and reallocation costs by optimizing the interval width for liquidity concentration.
Although liquidity concentration seems reasonable to increase the efficiency of using provided liquidity, this work shows that full-range liquidity positions (i.e.,~liquidity provisioning without liquidity concentration) are most profitable. We hope research and practice will leverage these findings to continue advancing the field of AMMs.

\section*{Acknowledgement}
This work was supported by funding from the topic Engineering Secure Systems of the Helmholtz Association (HGF) and by KASTEL Security Research Labs.
    \section*{Appendix} \label{sec:appendix}

\subsection{Formulation for Optimization Software}  \label{sec:gurobiformulation}
\renewcommand{\varcolorred}{varcolorblack} 
\renewcommand{\varcolorblue}{varcolorblack} 

The following presents a formulation that can be implemented and executed in optimization software.

\begin{subequations}\label{optproblemgurobi}
{\allowdisplaybreaks
\begin{align} 
    &\textbf{P}: \max\,
     \EX\left[\textcolor{\varcolorblue}{\rprice_T^2}\textcolor{\varcolorred}{\xhelp_T}+\textcolor{\varcolorred}{\yhelp_T} + \textcolor{\varcolorred}{\fees_T} \right]  \\[5pt]
    &\text{s.t.} \quad  \textcolor{\varcolorred}{\alpha} \geq 1 + \epsilon, \\
    &  \textcolor{\varcolorred}{\rprice_t} \leq \textcolor{\varcolorred}{\rpl_{t-1}} \Rightarrow \textcolor{\varcolorred}{z^l_t} = 1 \text{ else } 0   ,   \\
    &  \textcolor{\varcolorred}{\rprice_t} \geq \textcolor{\varcolorred}{\rpu_{t-1}} \Rightarrow \textcolor{\varcolorred}{z^u_t} = 1 \text{ else } 0   ,   \\
    &  \textcolor{\varcolorred}{z^m_t} = 1 - \textcolor{\varcolorred}{z^l_t} - \textcolor{\varcolorred}{z^u_t}    ,   \\
    & \textcolor{\varcolorred}{z_t} = \textcolor{\varcolorred}{z^l_t} + \textcolor{\varcolorred}{z^u_t} \\
    & \textcolor{\varcolorred}{\rpl_t} = \textcolor{\varcolorred}{z_t} \frac{1}{\textcolor{\varcolorred}{\alpha}}\textcolor{\varcolorblue}{\rprice_t} + \left(1-\textcolor{\varcolorred}{z_t}\right) \textcolor{\varcolorred}{\rpl_{t-1}}  ,   \\
    &  \textcolor{\varcolorred}{\rpu_t} = \textcolor{\varcolorred}{z_t} \textcolor{\varcolorred}{\alpha}\textcolor{\varcolorblue}{\rprice_t} + \left(1-\textcolor{\varcolorred}{z_t}\right) \textcolor{\varcolorred}{\rpu_{t-1}}  ,   \\
    &  \textcolor{\varcolorred}{\xhelp_t} = \textcolor{\varcolorred}{z^l_t}\textcolor{\varcolorred}{\liq_{t-1}} \left(\frac{1}{\textcolor{\varcolorred}{\rpl_{t-1}}}-\frac{1}{\textcolor{\varcolorred}{\rpu_{t-1}}}\right) 
    \nonumber  \\ & \quad \quad
    + \textcolor{\varcolorred}{z^m_t}\textcolor{\varcolorred}{\liq_{t-1}} \left(\frac{1}{\textcolor{\varcolorblue}{\rprice_t}}-\frac{1}{\textcolor{\varcolorred}{\rpu_{t-1}}}\right) , \\
    & \textcolor{\varcolorred}{\yhelp_t} = \textcolor{\varcolorred}{z^m_t} \textcolor{\varcolorred}{\liq_{t-1}} \left(\textcolor{\varcolorblue}{\rprice_t}-\textcolor{\varcolorred}{\rpl_{t-1}}\right) 
    \nonumber \\ &  \quad \quad
    + \textcolor{\varcolorred}{z^u_t} \textcolor{\varcolorred}{\liq_{t-1}} \left(\textcolor{\varcolorred}{\rpu_{t-1}}-\textcolor{\varcolorred}{\rpl_{t-1}} \right), \\
    &  \textcolor{\varcolorred}{z_t} \left(\textcolor{\varcolorblue}{\rprice_t}^2\textcolor{\varcolorred}{\xreal_t} + \textcolor{\varcolorred}{\yreal_t}\right) =  \textcolor{\varcolorred}{z_t} \left(\textcolor{\varcolorblue}{\rprice_t}^2\textcolor{\varcolorred}{\xhelp_t} + \textcolor{\varcolorred}{\yhelp_t} - \ctransact\left\vert\textcolor{\varcolorred}{\yreal_t} - \textcolor{\varcolorred}{\yhelp_t}\right\vert\right. 
    \nonumber \\ &  \qquad \qquad \qquad \qquad \quad 
    \left. - \gas + \textcolor{\varcolorred}{\fees_{t}}\right) , \\
    &  \textcolor{\varcolorred}{z_t} \left(\textcolor{\varcolorred}{\yreal_t}\right) = \textcolor{\varcolorred}{z_t} \textcolor{\varcolorblue}{\rprice_t^2} \textcolor{\varcolorred}{\xreal_t} , \\
    &  \left(1-\textcolor{\varcolorred}{z_t}\right)\textcolor{\varcolorred}{\xreal_t}=\left(1-\textcolor{\varcolorred}{z_t}\right)\textcolor{\varcolorred}{\xhelp_t},\\
    &  \left(1-\textcolor{\varcolorred}{z_t}\right)\textcolor{\varcolorred}{\yreal_t}=\left(1-\textcolor{\varcolorred}{z_t}\right)\textcolor{\varcolorred}{\yhelp_t},\\
    &  \textcolor{\varcolorred}{\liq_t} = \textcolor{\varcolorred}{z_t} \left(\frac{\textcolor{\varcolorred}{\alpha}}{\textcolor{\varcolorred}{\alpha}-1} \frac{\textcolor{\varcolorred}{\yreal_t}}{\textcolor{\varcolorblue}{\rprice_t}}\right) + \left(1-\textcolor{\varcolorred}{z_t}\right)\textcolor{\varcolorred}{\liq_{t-1}} , \\
    & \textcolor{\varcolorred}{\feesconst_t} = \textcolor{\varcolorred}{z^l_t} \frac{\textcolor{\varcolorblue}{\rprice_{t-1}^2}-\textcolor{\varcolorred}{\left(\rpl_{t-1}\right)^2}}{\textcolor{\varcolorblue}{\rprice_{t-1}^2}-\textcolor{\varcolorblue}{\rprice_{t}^2}}\overline{\feesconst} + \textcolor{\varcolorred}{z^m_t}\overline{\feesconst}
    \nonumber \\ &  \quad \quad
    +\textcolor{\varcolorred}{z^u_t} \frac{\textcolor{\varcolorred}{\left(\rpu_{t-1}\right)^2}-\textcolor{\varcolorblue}{\rprice_{t-1}^2}}{\textcolor{\varcolorblue}{\rprice_{t}^2}-\textcolor{\varcolorblue}{\rprice_{t-1}^2}}\overline{\feesconst}, \\
    &   \textcolor{\varcolorred}{\fees_t} = \textcolor{\varcolorred}{\fees_{t-1}} \left(1-\textcolor{\varcolorred}{z_{t-1}}\right) + \textcolor{\varcolorred}{\feesconst_t} \textcolor{\varcolorred}{\liq_{t-1}},   \\
    & \textcolor{\varcolorred}{z_t},\, \textcolor{\varcolorred}{z^l_t},\, \textcolor{\varcolorred}{z^m_t},\, \textcolor{\varcolorred}{z^u_t} \in \left\{0,\,1\right\} ,  \\
    &  \textcolor{\varcolorred}{\alpha},\, \textcolor{\varcolorred}{\rpl_t},\, \textcolor{\varcolorred}{\rpu_t},\,  \textcolor{\varcolorred}{\xhelp_t},\, \textcolor{\varcolorred}{\yhelp_t},\,\textcolor{\varcolorred}{\xreal_t},\, \textcolor{\varcolorred}{\yreal_t},\, \textcolor{\varcolorred}{\liq_t},\, \textcolor{\varcolorred}{\feesconst_t},\,  \textcolor{\varcolorred}{\fees_t} \geq 0, 
 \end{align}}
\end{subequations}

\subsection{Optimization Results and Runtime}
The optimization results for different time steps $T$ and sampling sizes $S$ are show in \cref{tab:descriptive_statistics_alpha}. 
The computations were executed on a MacBook Air with M1 chip and 8 GB of RAM. The runtime for our optimization is shown in \cref{tab:descriptive_statistics_runtime}. 

\begin{table}[ht]
    \centering
    \caption{Descriptive statistics of the optimal size $\size$ over 10 iterations for different time steps $T$ and sampling sizes $S$}
    \label{tab:descriptive_statistics_alpha}
    \setlength\tabcolsep{5pt}
    \begin{tabular}{|c|c|c|c|c|c|c|c|c|}
        \hline
        \multirow{1}{*}{\textbf{T}} & \multicolumn{4}{c|}{\boldmath{$5$}} & \multicolumn{4}{c|}{\boldmath{$10$}} \\
        \cline{1-9}
         \textbf{S} & \textbf{5} & \textbf{10} & \textbf{20} & \textbf{30} & \textbf{5} & \textbf{10} & \textbf{20} & \textbf{30} \\
        \hline
        Mean & 3.69 & 4.00 & 3.90 & 4.00 & 3.99 & 3.70 & 3.57 & 3.61 \\
        Median & 4.00 & 4.00 & 4.00 & 4.00 & 4.00 & 4.00 & 4.00 & 4.00 \\
        Min & 1.01 & 3.98 & 3.00 & 3.95 & 3.95 & 1.04 & 1.02 & 1.03 \\
        Max & 4.00 & 4.00 & 4.00 & 4.00 & 4.00 & 4.00 & 4.00 & 4.00 \\
        Std Dev & 0.94 & 0.01 & 0.32 & 0.01 & 0.02 & 0.93 & 0.93 & 0.95 \\
        \hline
    \end{tabular}
\end{table}

\begin{table}[ht]
    \centering
    \caption{Descriptive statistics of the runtime in seconds over 10 iterations for different time steps $T$ and sampling sizes $S$}
    \label{tab:descriptive_statistics_runtime}
    \setlength\tabcolsep{5pt}
    \begin{tabular}{|c|c|c|c|c|c|c|c|c|}
        \hline
        \multirow{1}{*}{\textbf{T}} & \multicolumn{4}{c|}{\boldmath{$5$}} & \multicolumn{4}{c|}{\boldmath{$10$}} \\
        \cline{1-9}
         \textbf{S} & \textbf{5} & \textbf{10} & \textbf{20} & \textbf{30} & \textbf{5} & \textbf{10} & \textbf{20} & \textbf{30} \\
        \hline
        Mean & 0.41 & 0.95 & 2.41 & 6.31 & 1.75 & 6.19 & 16.71 & 30.30 \\
        Median & 0.30 & 0.76 & 1.99 & 5.58 & 1.73 & 5.47 & 13.86 & 32.27 \\
        Min & 0.16 & 0.46 & 0.39 & 2.29 & 0.54 & 2.61 & 7.86 & 8.23 \\
        Max & 0.94 & 2.09 & 5.00 & 13.50 & 3.43 & 15.44 & 33.18 & 51.33 \\
        Std Dev & 0.27 & 0.54 & 1.47 & 3.30 & 0.94 & 3.84 & 9.17 & 12.27 \\
        \hline
    \end{tabular}
\end{table}

    \bibliographystyle{IEEEtran}
    \bibliography{01_content/ieee}

\end{document}